\documentclass{aip-cp}

\usepackage[numbers]{natbib}
\usepackage{rotating}
\usepackage{graphicx}

\newcommand{\bra}[1]{\langle #1|}
\newcommand{\ket}[1]{|#1\rangle}

\newcommand{\braket}[2]{\langle #1|#2\rangle}

\begin{document}

\title{Quantum Measurement and Weak Values in Entropic Dynamics\footnote{%
 Presented at MaxEnt 2016, the 36th International Workshop on Bayesian Inference and Maximum Entropy Methods in Science and Engineering (July 10-15, 2016, Ghent, Belgium).}}

\author[aff1]{Kevin Vanslette\corref{cor1}}
\author[aff1]{Ariel Caticha\corref{cor2}}

\affil[aff1]{Department of Physics, University at Albany - SUNY, Albany, NY 12222, USA}
\corresp[cor1]{kvanslette@albany.edu}
\corresp[cor2]{arielcaticha@gmail.com}

\maketitle

\begin{abstract}
The problem of measurement in quantum mechanics is studied within the Entropic Dynamics framework. We discuss von Neumann and Weak measurements, wavefunction collapse, and Weak Values as examples of bayesian and entropic inference.
\end{abstract}
\section{INTRODUCTION}
Since its formulation, Quantum Mechanics (QM) has always peaked the interest of
individuals with its inherent mystery and its seemingly paradoxical, but
experimentally reproducible, results. At the heart of this mystery is the
superposition principle and the measurement problem. The situation has been
nicely described by Wallace \cite{Wallace}: \textquotedblleft Solutions of
the measurement problem are often called `interpretations of QM', the idea
being that all such `interpretations' agree on the formalism and thus the
experimental predictions", and thus the interpretation of QM in Entropic
Dynamics \cite{ED,book}\emph{\ }will inevitably
be discussed in the context of the measurement problem.

ED differs from most physical theories. In most approaches to QM one starts
with the formalism and the interpretation is appended to it, almost as an
afterthought. In ED one starts with the interpretation, that is, one
specifies what the ontic elements of the theory are, and only then one
develops a formalism appropriate to describe, predict and control those ontic elements. For instance, to
derive non-relativistic QM one starts from the assumption that particles
have definite yet unknown positions and then one proceeds to derive ED as a
non-dissipative diffusion.

The solution to the measurement problem should address two problems: one is
the problem of \emph{definite outcomes} (when and how does the wavefunction collapses), the other is the problem of the \emph{preferred basis},
which is also known as the \emph{basis degeneracy problem}. (For a review see 
\cite{Schlosshauer}.) Both problems as well as the interpretation
of QM in ED will be addressed in the body of this paper.

Entropic Dynamics (ED) derives laws of physics as an example entropic inference.\footnote{%
The principle of maximum entropy as a method for inference can be traced to
the pioneering work of E. T. Jaynes \cite{Jaynes 1957}\cite{Jaynes 2003}.
For a pedagogical overview of Bayesian and entropic inference and for
further references see \cite{book}.} This view is extremely
constraining. For example, there is no room for \textquotedblleft quantum
probabilities\textquotedblright\ --- probabilities are neither classical nor
quantum, they are tools for reasoning that have universal applicability.
Related to this is the fact that once one claims that the wavefunction is
an epistemic object (\emph{i.e.}, $|\Psi |^{2}$ is a probability) then its
time evolution --- the updating of $\Psi $ --- is not at all arbitrary. The
laws of dynamics must be dictated by the usual rules of inference.

There is one feature of the unified Bayesian/entropic inference approach
that turns out to be particularly relevant to the foundations of quantum
theory. The issue in question is von Neumann's proposal of two modes of
evolution for the wavefunction, either continuous and unitary as given by
the Schr\"{o}dinger equation, or discontinuous and stochastic as described
by the projection postulate -- the collapse of the wavefunction. Once one
adopts the broader inference scheme that recognizes the consistency of
Bayesian updating with entropic updating \cite{Giffin} the
apparent conflict between the two modes of wavefunction evolution
disappears: The wavefunction is an epistemic object meant to be updated on
the basis of new information. When the system is isolated the evolution is obtained by a continuous entropic updating which leads to a unitary evolution described by the Schr\"{o}dinger equation. When, in a measurement, the information is in the form of discrete data the Bayesian update is discontinuous and the wavefunction collapses. Both forms of updating coexist harmoniously within the framework of ED \cite{Johnson}.

In the von Neumann measurement scheme, the macroscopic measurement device is
treated quantum mechanically \cite{Neumann}. The initial state of the
measuring device is generally given as $|r\rangle $ in indicating the device
is in the \textquotedblleft ready\textquotedblright\ or \textquotedblleft
reference\textquotedblright\ state. An interaction between the system, in an
initial state  $\ket{\Psi}=\sum \alpha _{n}\ket{a_{n}}$ and the
measuring device leads to a state in which the system becomes entangled with
the device's \textquotedblleft pointer\textquotedblright\ variable,
\begin{eqnarray}
\Big(\sum_n \alpha_n\ket{a_n}\Big)\ket{r}\stackrel{t}{\longrightarrow}\sum_n\alpha'_{nn}\ket{a_n}\ket{b_n}\label{vN a}.
\end{eqnarray}
The standard von Neumann measurement scheme involves interactions such that
the system and measurement device evolve into a special entagled state called a \textquotedblleft
biorthogonal\textquotedblright\ state (\emph{i.e.}, there are no cross terms
in eq.(\ref{vN a}), $\alpha _{mn}^{\prime }=0$ for $m\neq n$) \cite%
{Schlosshauer}. Thus, a measurement that finds the pointer variable in state 
$|b_{n}\rangle $ seemingly allows the observer to infer that the system is in state $%
|a_{n}\rangle $. 
However, the right hand side of eq.(\ref{vN a}) can be expanded in some other basis. There is no ontological matter of fact about wihich outcome has been obtained. Thus we have problems: not only we have a never observed phenomenon of macroscopic entanglement but we also have a \emph{problem of no definite outcomes} also called the \emph{preferred basis problem} or the \emph{basis degeneracy problem} \cite{Schlosshauer}\cite{Steeb}.

In ED the collapse of the wave function involves a straightforward application of Bayes theorem to find $%
P(a_{n}|x_{m})$, the probability the system is in state $|a_{n}\rangle $
given a detection of the pointer variable in state $|x_{m}\rangle $. In ED
it is possible and convenient to introduce \textquotedblleft
observables\textquotedblright\ other than position (\emph{e.g.}, momentum,
energy, and so on) but these are not attributes of the system, they are
attributes of their probability distributions. 
While positions are
\textquotedblleft ontic\textquotedblright\ elements, these other observables
are purely epistemic; they are properties of the wave function, not of the system \cite{Johnson}. These ideas can be
pushed to an extreme when discussing the notion of the Weak Value \cite{aav}-\cite{Wiseman} of an operator $A$ in
which the system is prepared in an initial state $\ket{\Psi}$ and
post-selected in state $|\Psi ^{\prime }\rangle $, 
\begin{equation}
A_{w}=\frac{\langle \Psi ^{\prime }|A|\Psi \rangle }{\langle \Psi ^{\prime
}|\Psi \rangle }~.
\end{equation}%
Weak values are complex and therefore strictly they cannot be observed, and yet they may
still be \emph{inferred}. From the perspective of an inference framework such as ED
the common reference to `observables' is misguided. It should be replaced by
Bell's term `beables' \cite{Bell 1993} for ontic elements such as particle
positions, and perhaps `inferables' for those epistemic elements associated
to probability distributions.\footnote{%
Clearly, beables are inferables too.}

After a brief review of ED and of the simplest or direct type of measurement 
\cite{Johnson}, we discuss the less direct types of measurement ---
the von Neumann and weak measurements --- in which information about the
system is inferred by observation of another system, the pointer device, to
which it has been suitably correlated. Finally we discuss Weak Values in the
context of ED.
\section{Entropic dynamics --- a brief review}

Here we give a brief overview of ED emphasizing those ideas relevant to the
measurement problem. (For a review with more extended references see \cite%
{ED}.) The system consists of $N$ particles living in a flat
Euclidean space $\mathbf{X}$ with metric $\delta _{ab}$. The positions of
the particles have \emph{definite} values which are however \emph{unknown}.
We already see that ED differs from the Copenhagen interpretation: in the
latter, positions become definite only as a result of the measurement
process.

Positions are denoted $x_{n}^{a}$ where $n=1,\ldots, N$ denotes the particle
and $a=1,2,3$ denotes the spatial coordinates. The microstate $x^{A}$, where $%
A=(n,a)$, is a point in the $3N$ dimensional configuration space. 

The objective is to predict the particle positions and their motion. The
first assumption is that particles follow continuous trajectories so that
the motion can be analyzed as a sequence of short steps. We use the method
of maximum entropy to find the probability $P(x^{\prime }|x)$ that the
particles take infinitesimally short steps from $x_{n}^{a}$ to $%
x_{n}^{\prime a}=x_{n}^{a}+\Delta x_{n}^{a}$. The information that the steps
are infinitesimal is given by $N$ constraints,  
\begin{equation}
\langle \Delta x_{n}^{a}\Delta x_{n}^{b}\rangle \delta _{ab}=\kappa
_{n}~,\qquad (n=1,\ldots, N)~~  \label{kappa n}
\end{equation}%
(eventually we take the limit $\kappa _{n}\rightarrow 0$ imposing continuous motion). To introduce
correlations we impose one additional constraint, 
\begin{equation}
\langle \Delta x^{A}\rangle \partial _{A}\phi
=\sum\limits_{n,a}\left\langle \Delta x_{n}^{a}\right\rangle \frac{%
\partial \phi }{\partial x_{n}^{a}}=\kappa ^{\prime }~,  \label{kappa prime}
\end{equation}%
where $\partial _{A}=\partial /\partial x^{A}=\partial /\partial x_{n}^{a}$, 
$\kappa ^{\prime }$ is another small constant, and $\phi $ is the
\textquotedblleft drift\textquotedblright\ potential. It is this single
constraint acting on the $3N$ dimensional configuration space that leads to
quantum effects such as interference and entanglement. The physical nature of the 
$\phi $ potential need not concern us here. We merely postulate its
existence and note $\phi $ is closely related to the phase of the wave
function.\footnote{%
Since $\phi $ affects the motion of particles it plays the role of a pilot
wave or an electromagnetic field. Indeed, $\phi $ is \emph{as real as} the
vector potential $A^{a}$. The close relation between them is the gauge
symmetry (see \cite{ED}).} The result of maximizing entropy with
an appropriate choice of Lagrange multipliers leads to 
\begin{equation}
P(x^{\prime }|x)=\frac{1}{Z}\exp [-\sum_{n,a}(\frac{m_{n}}{2\eta \Delta t}%
\left( \,\Delta x_{n}^{a}-\langle \Delta x_{n}^{a}\rangle \right) \left(
\,\Delta x_{n}^{a}-\langle \Delta x_{n}^{a}\rangle \right) ]~,
\label{Prob xp/x a}
\end{equation}%
where $Z$ is a normalization constant, $\Delta t$ is the time interval
between $x$ and $x^{\prime }$, the $m_{n}$'s are particle-specific constants
called \textquotedblleft masses\textquotedblright ,\ and $\eta $ is a
constant that fixes the units of time relative to those of length and mass 
\cite{ED}. Once we have $P(x^{\prime }|x)$ for an infinitesimal
step we can ask how a probability distribution $\rho(x,t)$ evolves from one step to the next step $\rho(x',t')$ by considering,
\begin{equation}
\rho(x',t')=\int P(x^{\prime }|x)\rho(x,t).\label{step}
\end{equation}
 Iterating this process for very short steps allows one to find the evolution of $\rho (x,t)$ to be given by the Fokker-Planck equation (an explication can be found in \cite{book}), 
\begin{equation}
\partial _{t}\rho =-\partial _{A}\left( \rho v^{A}\right),   \label{FP b}
\end{equation}%
where $v^{A}$ is the velocity of the probability flow in configuration space
or \emph{current velocity}, 
\begin{equation}
v^{A}=m^{AB}\partial _{B}\Phi \quad \mbox{where}\quad m^{AA^{\prime
}}=m_{n}^{-1}\delta ^{aa^{\prime }}\delta _{nn^{\prime }}\quad \mbox{and}%
\quad \Phi =\eta \phi -\eta \log \rho ^{1/2}~.  \label{curr}
\end{equation}%
Equation (\ref{FP b}) describes a standard diffusion. To obtain a
\textquotedblleft non-dissipative\textquotedblright\ dynamics the choice of
drift potential $\phi $ must be revised after each step $\Delta t$, which
means that the drift potential, or equivalently, the \textquotedblleft
phase\textquotedblright\ $\Phi $, is promoted to a dynamical degree of
freedom. The correct updating, $\Phi \rightarrow \Phi +\delta \Phi $,
follows from requiring that a certain functional $\tilde{H}[\rho ,\Phi ]$ be
conserved. The result is that the coupled evolution of $\rho $ and $\Phi $
is given by a conjugate pair of Hamilton's equations, 
\begin{equation}
\partial _{t}\rho =\frac{\delta \tilde{H}}{\delta \Phi }\quad \mbox{and}%
\quad \partial _{t}\Phi =-\frac{\delta \tilde{H}}{\delta \rho }~.
\label{Hamilton}
\end{equation}%
The \textquotedblleft ensemble\ Hamiltonian\textquotedblright\ $\tilde{H}$
is chosen so that the first equation above reproduces (\ref{FP b}) in which case
the second equation in (\ref{Hamilton}) becomes a Hamilton-Jacobi equation. A
more complete specification of $\tilde{H}$ suggested by information
geometry, 
\begin{equation}
\tilde{H}[\rho ,\Phi ]=\int dx\,\left[ \frac{1}{2}\rho m^{AB}\partial
_{A}\Phi \partial _{B}\Phi +\rho V+\xi m^{AB}\frac{1}{\rho }\partial
_{A}\rho \partial _{B}\rho \right] ~,  \label{Hamiltonian}
\end{equation}%
includes the \textquotedblleft kinetic\textquotedblright\ term that
reproduces (\ref{FP b}), a potential $V(x)$, and a \textquotedblleft
quantum\textquotedblright\ potential. The parameter $\xi =\hbar ^{2}/8$\
defines the value of Planck's constant $\hbar $. The formulation of ED is
now complete: combining $\rho $ and $\Phi $ into a single complex function, $%
\Psi =\rho ^{1/2}\exp (i\Phi /\hbar )$ with $\hbar =(8\xi )^{1/2}$ allows us
to rewrite (\ref{Hamilton}) as a \emph{linear }Schr\"{o}dinger equation,%
\begin{equation}
i\hbar \frac{\partial \Psi }{\partial t}=-\sum_{n}\frac{\hbar ^{2}}{2m_{n}}%
\bigtriangledown^2_n\Psi +V\Psi ~.  \label{sch c}
\end{equation}

At this point we can adopt the standard Hilbert space formalism to represent
the epistemic state $\Psi (x)$ as a vector, 
\begin{equation}
|\Psi \rangle =\int dx\,\Psi (x)|x\rangle \quad \mbox{with\quad }\Psi
(x)=\langle x|\Psi \rangle ~,
\end{equation}%
We can express $|\Psi \rangle $ in bases other than the position basis $%
\{|x\rangle \}$ and we can write the Schr\"{o}dinger equation as a
manifestly unitary evolution, $|\Psi (t)\rangle =U(t,t_{0})|\Psi
(t_{0})\rangle $ for some suitably chosen $U(t,t_{0})$.

The particles have definite positions but their Brownian paths are not
smooth; there is no momentum associated to the trajectories. One can define
the usual quantum momentum --- the operator that generates translations of
the wavefunction. In general it does not have definite values. Its values,
which are inferred by the process of measurement and detection, are not an
\textquotedblleft ontic\textquotedblright\ attribute of the particles but an
\textquotedblleft epistemic\textquotedblright\ attribute of the wave
function. 

The simplest type of quantum measurement, studied in \cite{Johnson},
consists of the direct detection of the particle's position. Since the wavefunction is purely epistemic, the collapse
problem does not arise. The \textquotedblleft
collapse\textquotedblright\ is nothing but a Bayesian update from a prior to
a posterior and in general the preferred basis problem for a single orthonormal vector does not apply \cite{Schlosshauer,Steeb}. In this context, a question of interest in ED is how do observables other
than position arise was also answered in \cite{Johnson}. Consider, for simplicity, a particle that lives on a
discrete lattice, so that if the state is 
\begin{equation}
|\Psi \rangle =\sum\nolimits_{i}c_{i}|x_{i}\rangle \quad \mbox{then}\quad 
\mbox{Prob}(x_{i})=p_{i}=|\langle x_{i}|\Psi \rangle |^{2}=|c_{i}|^{2}~.
\end{equation}%
In a more \textquotedblleft complicated\textquotedblright\ measurement the
particle is subject to additional interactions right before reaching the
position detector. Such a complicated setup $\mathcal{A}$ is described by a
particular \emph{unitary} evolution $\hat{U}_{A}$. The particle will evolve to the position $|x_{i}\rangle $ with certainty provided it was
initially in a state $|a_{i}\rangle $ such that $\hat{U}_{A}|a_{i}\rangle
=|x_{i}\rangle $. Since the set $\{|x_{i}\rangle \}$ is orthonormal and
complete, the set $\{|a_{i}\rangle \}$ is also orthonormal and complete. To
figure out the effect of $\mathcal{A}$ on some generic initial state vector $%
|\Psi \rangle $, expand as $|\Psi \rangle
=\sum\nolimits_{i}c_{i}|a_{i}\rangle \ $where$\ c_{i}=\langle a_{i}|\Psi
\rangle $. Then the state $|\Psi \rangle $ evolves according to $\hat{U}_{A}$
into the new state at a later time $t'$,
\begin{equation}
\hat{U}_{A}|\Psi \rangle =\sum\nolimits_{i}c_{i}\hat{U}_{A}|a_{i},t\rangle
=\sum\nolimits_{i}c_{i}|x_{i},t'\rangle \ ,\label{unitary}
\end{equation}%
which, invoking the Born rule for \emph{position} measurements, implies that
the probability of finding the particle at the position $x_{i}$ is $%
p_{i}=|c_{i}|^{2}=|\langle x_{i}|\Psi(t') \rangle |^{2}=|\langle a_{i}|\Psi(t) \rangle |^{2}$. 

From a physics perspective there is nothing more to say but we can adopt 
\emph{a different language}: we can \emph{say} that the particle has been
\textquotedblleft measured\textquotedblright\ \emph{as if} it had earlier been in
the state $|a_{i}\rangle $. Thus, the setup $\mathcal{A}$ is a complicated device that in principle
\textquotedblleft measures\textquotedblright\ all operators of the form $%
\hat{A}=\sum\nolimits_{i}\lambda _{i}|a_{i}\rangle \langle a_{i}|$ where
the eigenvalues $\lambda _{i}$ are \emph{arbitrary} scalars. Note that there
is no implication that the particle previously had or now currently has the
value $\lambda _{i}$ --- this is just a figure of speech. Von Neumann measurements are explored briefly and Bayes' Rule is used as an instance of ``collapse" in amplification processes \cite{Johnson}. 
\section{MEASUREMENTS AND DETECTIONS IN ENTROPIC DYNAMICS}
In this section, the von Neumann and weak measurement schemes are naturally implemented into ED. We will attempt to be particularly careful in our word choice such that a \emph{measurement} in ED involves first an \emph{interaction} with an apparatus or with another auxiliary system with which it becomes entangled. This step does not collapse the wave function. Then there is the actual \emph{detection}, which is the step that collapses or localizes the distribution to the value detected by the observer.
\subsection{Detection in ED: The ``Collapse" of the Wavefunction}

Here we present ED's solution to the problem of \emph{definite outcomes} and discuss the ``collapse" of the wavefunction.
Entropic Dynamics is a theory of inference and the rules for probability updating are precisely the same as in any other inference problem. There is no such thing as quantum probabilities.  
The problem of ``physical"  collapse is never encountered in ED for the same reason that it is not encountered in epistemic classical probability theory. No-one asks how the probability distribution of a die role collapses during measurement; this just follows from the inductive logic expressed by Bayes' Rule. 

When a particle with wave function $\Psi$ is detected at $x_D$ with certainty the prior probability $\rho(x) = |\braket{x}{\Psi}|^2$ is updated to $P'(x) =\delta (x-x_D)$. A more realistic detector is represented (for the purpose of inference) by a likelihood function $q(D|x)$ that gives the probability of a detection $D$ when the particle is at $x$. Then the probability that the particle was at $x$ given the detection event $D$ is given by Bayes' rule,
\begin{eqnarray}
P'(x)=q(x|D)=\frac{\rho(x)q(D|x)}{q(D)}.\label{detect}
\end{eqnarray}
An idealized device detects the presence of a particle in an infinitesimally small region centered around $x_D$ so the likelihood function in this case is $q(D|x)=\delta (x-x_D)$ and indeed $P'(x)=\delta(x-x_{D})$.
\subsection{Von Neumann and Weak Measurements in Entropic Dynamics}
As mentioned in the introduction, a von Neumann measurement scheme \cite{Neumann} is one in which a detection of the pointer variable in state $\ket{b_n}$ gives $\ket{a_n}$ with certainty and it potentially suffers from the \emph{preferred basis problem} or equivalently the \emph{basis degeneracy problem} \cite{Schlosshauer, Steeb}. We circumvent this problem by using Bayes Theorem and a specific unitary evolution to entangle the system and measurement device via the Weak Measurement scheme \cite{aav}. In ED the only positions are beables, all other ``observables" must be inferred from detections of position and strictly are properties of the wavefunction or probability distribution. In this sense, quantities such as momentum or energy are not ontic (in ED) but rather are seen to be convenient parameters for expressing the probability distribution in question, that is, they are epistemic quantities verifying \cite{Johnson}. For instance, an expectation value can be a useful parameter for expressing our knowledge of the state of a system but in general it is a property belonging to the probability distribution rather than being something persistently expressed through the system as an ontic variable is. The simplest pointer measurement device we can imagine in this framework uses the position of a single test particle $x$ as its pointer variable. The construction of a macroscopic pointer measurement device is an exercise in the central limit theory for a many particle system \cite{book,Demme}. 

To avoid clutter we will neglect normalization factors until we reach the level of probability distributions near the end of the calculation. An ideal pointer measuring device in ED is one in which a test particle has a definite initial state,
\begin{eqnarray}
\ket{\Phi_{i}}_{ideal}=\int \delta(x-x_{i})\ket{x}\,dx,
\end{eqnarray}
in position space and we will let $x_i=0$ for simplicity.  We consider a more general case in which the test particle is in the initial state,
\begin{eqnarray}
\ket{\Phi_{i}}=\int e^{-\frac{x^2}{4\bigtriangleup^2}}\ket{x}\,dx,
\end{eqnarray}
which reproduces the ideal measurement device when $\bigtriangleup\rightarrow0$ after normalization.  Using the completeness relationship $1=\int dp \ket{p}\bra{p}$, the state of the test particle can be represented in momentum space as,
\begin{eqnarray}
\ket{\Phi_{i}}=\int e^{-\frac{x^2}{4\bigtriangleup^2}}\int  \ket{p}\braket{p}{x}\,dxdp=\int e^{-\frac{x^2}{4\bigtriangleup^2}}\int  e^{-ip\cdot x}\,dx\ket{p}dp=\int e^{-\bigtriangleup^2p^2} \ket{p}\,dp.
\end{eqnarray}
The quantum system to be measured with the pointer device is a preselected or prepared superposition state of the eigenvectors of $\hat{A}$,
\begin{eqnarray}
\ket{\Psi_{i}}=\sum_{n}\alpha_n\ket{A=a_n}.
\end{eqnarray}
We will consider a situation in which the measuring device is coupled to the system to be measured by a strong coupling or interaction Hamiltonian,
\begin{eqnarray}
\hat{H}=-g(t)\hat{p}\hat{A},\label{H}
\end{eqnarray}
where $\hat{p}$ is the canonical conjugate of the pointer variable and thereby generates translations in the pointer variable $\hat{x}$ and $g(t)$ is a function with compact support near the time of measurement and integrates to unity \cite{aav}. We can assume that the coupling Hamiltonian will dominate over the full Hamiltonian for the, assumed small, period of measurement. The time evolution of our entangled system is,
\begin{eqnarray}
\ket{\Psi_{f},\Phi_{f}}=U\ket{\Psi_{i}}\ket{\Phi_{i}}=e^{-i\int \hat{H} dt}\ket{\Psi_{i}}\ket{\Phi_{i}}\nonumber
\end{eqnarray}
\begin{eqnarray}
=\sum_{n}\alpha_n\int e^{-\bigtriangleup^2p^2}e^{i\hat{p}\hat{A}}\ket{A=a_n}\ket{p}\,dp
=\sum_{n}\alpha_n \int e^{-\bigtriangleup^2p^2}e^{ipa_n} \ket{A=a_n}\ket{p}\,dp\label{ent},
\end{eqnarray}
which, in the position space representation, is 
\begin{eqnarray}
=\sum_{n}\alpha_n\int e^{-\bigtriangleup^2p^2}e^{-ip(x-a_n)}\ket{A=a_n}\ket{x}\,dp\,dx=\sum_{n}\alpha_n\int e^{-\frac{(x-a_n)^2}{4\bigtriangleup^2}} \ket{A=a_n}\ket{x}\,dx,\label{24}
\end{eqnarray}
This is a superposition of potentially overlapping Gaussian distributions having peaks at the eigenvalues of $A$. When the Gaussian distributions overlap we have a so-called ``weak measurement" \cite{aav}; when the Gaussian distributions are neatly resolved we have a ``strong'' or von Neumann measurement. The joint probability of the system,
\begin{eqnarray}
P_f(a_n,x)=|\braket{a_n,x}{\Psi_{f},\Phi_{f}}|^2=\frac{|\alpha_n|^2e^{-\frac{(x-a_n)^2}{2\bigtriangleup^2}}}{Z}  .
\end{eqnarray}
In the present case, we may infer the probability the system of interest is accurately \emph{describable} by a particular eigenvector $\ket{a_n}$ by making detections of the position of the pointer particle following notions from the ``Detection in ED:..." section and \cite{Johnson}. This is given by the conditional probability,
\begin{eqnarray}
P(a_n|x')=\frac{P(a_n,x')}{P_f(x')}=\frac{|\alpha_n|^2  e^{-\frac{(x'-a_n)^2}{2\bigtriangleup^2}}}{\sum_{m}|\alpha_m|^2  e^{-\frac{(x'-a_m)^2}{2\bigtriangleup^2}}},\label{22a}
\end{eqnarray}
Thus the detection of the test particle at $x'$ allows us to infer the probability that the system is in a state with eigenvalue $a_n$. Bayes Rule leads to a partial collapse because the initial state of the pointer variable is itself uncertain. In the limit that $\bigtriangleup\rightarrow 0$ for $P_f(x)$ (after normalization) we have an ideal (von Neumann) measurement. The measurement always produces one of its eigenstates with probability $|\alpha_n|^2$, and from a detection of the position $x'$ we can infer the eigenstate $\ket{a_n}$ with probability 1 using (\ref{22a}). Because position is ontic in ED it always plays the role of the preferred basis, which eliminates the basis degeneracy problem.
%

So far we have taken into account the uncertainty in the preparation of the pointer variable but we have assumed that the pointer variable $x$ of the test particle has been measured precisely. If the detection of $x$ is noisy then there is a second source of uncertainty quantified by $P'(x')$ given by equation (\ref{detect}).  The probability of the value $a_n$ is
\begin{eqnarray}
P'(a_n)=\int dx'\,P(a_n|x')P'(x').
\end{eqnarray}
By compounding a unitary detector $U_v$ from (\ref{unitary}), with the weak measurement devices above one can map a non-position pointer variable $\hat{v}$ to a position pointer variable $\hat{x}$ which in principle allows for a more robust set of operators $\hat{B}$ to be measured. We may then consider coupling Hamiltonians $H_c=g(t)\hat{B}\hat{v}_{conj}$, where $\hat{v}_{conj}$ is the Fourier conjugate to $\hat{v}$, and before detection apply $U_v=\sum_i\ket{x_i}\bra{v_i}$ to put the pointer variable in the position basis.
\section{Weak Values in Entropic Dynamics}
The Weak Value $A_w\equiv \bra{\Psi'}\hat{A}\ket{\Psi}/\braket{\Psi'}{\Psi}$ of a Hermitian operator $\hat{A}$ was first introduced by Aharonov, Albert, and Vaidman \cite{aav} (AAV) in 1989 as an interesting application of a weak measurement which gives nonintuitive results (recent review \cite{Dressel}). What is particularly significant in AAV's paper \cite{aav} is not that they defined an odd quantity associated tor $\hat{A}$, but rather they found, after a series of approximations (see \cite{duck}) a way in which $A_w$ could be ``measured.'' A Weak Value is a complex number which may lie outside the set of eigenvalues of $\hat{A}$ when $\braket{\Psi'}{\Psi}$ is sufficiently small. Due to this, the interpretation of Weak Values has had a ``colorful history" \cite{Dressel}, much of which can be summarized by the question, "Are Weak Values ontic properties of a particle?".

In ED, variables other than position either are inferred from position, are useful parameters in the position probability distribution of the particle, or may be a particularly convenient basis for representing the position space probability distribution $\rho(x)=|\Psi(x)|^2$ due to the completeness relations. From this perspective it is clear that inferring $A_w$ does not indicate that the particle is ontically expressing $A_w$. Consider the final state of an entangled detector from (\ref{ent}), given we consider the state is post-selected into $\ket{\Psi'}=\sum \alpha_n'\ket{A=a_n}$ :
\begin{eqnarray}
\ket{\Phi_f}=\bra{\Psi'}U\ket{\Psi}\ket{\Phi_i}=\bra{\Psi'}e^{-i\int \hat{H} dt}\ket{\Psi}\ket{\Phi_{i}}
\approx (\braket{\Psi'}{\Psi}+i\hat{p}\bra{\Psi'}\hat{A}\ket{\Psi}+...)\ket{\Phi_i}\nonumber\\
=\braket{\Psi'}{\Psi}(1+i\hat{p}A_w+...)\ket{\Phi_i} \approx \braket{\Psi'}{\Psi}\int dp e^{ipA_w}e^{-\bigtriangleup^2p^2}\ket{p}
\end{eqnarray}
which in the position space representation gives,
\begin{eqnarray}
\ket{\Phi_f}\approx \braket{\Psi'}{\Psi}\int dx \,\exp\Big(-\frac{(x-A_w)^2}{4\bigtriangleup^2}\Big)\ket{x},
\end{eqnarray}
given the set up configuration is in the desired range of validity \cite{duck,Dressel}. In the ED framework we are interested in the probability distribution of $x$ post-selected (just like one may preselect a quantum state) into $\Psi'$,
\begin{eqnarray}
P(x|\Psi')=P(x)=\frac{1}{Z}\exp\Big(-\frac{(x-\Re[A_w])^2}{2\bigtriangleup^2}\Big),
\end{eqnarray}
where $A_w$ is taken as a feature of the probability distribution of $x$.  Because $A_w$ appears as a parameter in the probability distribution of $x$, we may consider $P(x)=P(x|A_w)$ and invert the problem to ask, ``what is the probability the parameter $A_w$ has a certain value given a detection of the pointer particle at $x'$" -- that is we may use $P(A_w|\{x\})\propto\exp\Big(-\frac{(\Re[A_w]-\overline{x})^2}{2\bigtriangleup^2/N}\Big)$ and the parameter estimation scheme in \cite{book} to find $A_w$ in agreement with \cite{aav}. 

There are several potentially interesting Weak Values but in particular consider the operator $\hat{A}=\ket{x}\bra{x}$ post selected in a momentum state,
\begin{eqnarray}
A_{w}=\frac{\braket{p}{x}\braket{x}{\Psi}}{\braket{p}{\Psi}}\stackrel{p=0}{\rightarrow}  k\Psi(x),
\end{eqnarray}
which is proportional to the full wavefunction and $k$ is a constant that will be removed after normalization \cite{Lundeen1}. Lundeen \emph{et al} show that the real and imaginary parts of $\Psi(x)$ are proportional to the position and momentum shifts of the pointer variable and claim they are measuring the wavefunction. From the ED perspective it is more appropriate to state that the value of the wavefunction at each $x$ is being inferred. 
If the value of $\Psi(x)$ is inferred with certainty then it is possible to solve for the phase of the wavefunction (up to an additive constant $2\pi n$) and also infer the values of the drift potential and its derivatives. This provides a link to Wiseman's use of Weak Values to measure the probability current in Bohmian Mechanics \cite{Wiseman}. 
Especially in cases as exotic as the above, ED takes the standpoint that Weak Values and quantities other than position (energy, momentum, etc.) are best considered as epistemic ``inferables" rather than ontic beables or observables.
 
\section{CONCLUSIONS}
We have discussed the von Neumann and weak measurement schemes in the framework
of Entropic Dynamics and have offered solutions to the problems of \emph{definite values} and of the \emph{preferred basis}. 
From the perspective of ED quantum mechanics is a framework for updating probabilities, for processing information. When the quantum system is left undisturbed it evolves smoothly and the appropriate tool for updating is the Schr\"{o}dinger equation; when the system is coupled to a measuring device the appropriate tool to handle information in the form of data is Bayes theorem. There is no conflict: information of different types is handled differently. 

The fact that position plays a privileged role ---  positions are "beables" --- provides us with a natural pointer variable. In ED quantities normally called "observables" such as energy and momentum, and the quantities called "weak values" are not ontic properties of the system. They are epistemic properties of the wave function. None of these quantities are beables and they cannot be observed directly. Instead they can be inferred by appropriate position measurements either of the system itself or of another system with which it is suitably entangled. Such quantities should more appropriately be referred as "inferables.''
It is the explicit recognition of the ontic nature of positions vs the epistemic nature of other inferables that allows one to circumvent the paradoxes and confusion that have surrounded the problem of measurement. 

%

\section{ACKNOWLEDGMENTS}
We would to like the reviewers and also: M. Abedi, D. Barolomeo, N. Carrara, A. Fernandes, T. Gai, S. Ipek, K. Knuth, S. Nawaz, P. Pessoa, J. Tian, J. Walsh, and A. Yousefi for their many discussions on Entropy, Inference, and Quantum Mechanics.

%

\end{document}